NANO EXPRESS                                                             Open Access

# Electrical transport properties of small diameter single-walled carbon nanotubes aligned on ST-cut quartz substrates

Tohru Watanabe[1], El-Hadi S Sadki[2*], Takahide Yamaguchi[1] and Yoshihiko Takano[1]

**Abstract**

A method is introduced to isolate and measure the electrical transport properties of individual single-walled carbon nanotubes (SWNTs) aligned on an ST-cut quartz, from room temperature down to 2 K. The diameter and chirality of the measured SWNTs are accurately defined from Raman spectroscopy and atomic force microscopy (AFM). A significant up-shift in the G-band of the resonance Raman spectra of the SWNTs is observed, which increases with increasing SWNTs diameter, and indicates a strong interaction with the quartz substrate. A semiconducting SWNT, with diameter 0.84 nm, shows Tomonaga-Luttinger liquid and Coulomb blockade behaviors at low temperatures. Another semiconducting SWNT, with a thinner diameter of 0.68 nm, exhibits a transition from the semiconducting state to an insulating state at low temperatures. These results elucidate some of the electrical properties of SWNTs in this unique configuration and help pave the way towards prospective device applications.

## Background

Single-walled carbon nanotubes (SWNTs), with their miniature size, low structural defects, and various other superior properties [1-4], are very attractive nanomaterials as basis for future electronic devices [5-7]. However, there are still many technical obstacles towards the realization of SWNT-based devices, such as the difficulty of their positioning on a substrate, as well as the lack of control of their chirality, which eventually defines their electronic properties. Furthermore, synthesized SWNTs by chemical vapor deposition (CVD) on a substrate are usually short (around 10 μm) and randomly dispersed, which makes it difficult for device fabrication. Recently, it has been reported that arrays of long (hundreds of microns) and horizontally highly aligned SWNTs could be synthesized on some single crystal substrates, such as ST-cut quartz [8] and sapphire [9]. This is an important breakthrough, as the length of the synthesized SWNTs, and their high alignment, makes their electrical characterization and device fabrication much more accessible than ever before. Indeed, a field-effect transistor (FET) has been demonstrated using aligned SWNT arrays on an ST-cut quartz substrate [8]. It is also noted that the latest Raman and photoluminescence data suggest that these SWNTs have predominantly semiconducting properties [10,11]. However, and despite a lot of research work on SWNT array on ST-cut quartz [10,12,13], no data has been reported so far on the electrical properties or device fabrication of a single isolated SWNT on these substrates, except after their transfer onto silicon substrates [7]. We believe that this is important in order to understand the underlying physics of the SWNTs in this unique configuration, which is crucial for any prospective device applications. Furthermore, it has been reported recently that the aligned SWNTs on ST-cut quartz substrates are in strong interaction with the substrate [14,15], and the understanding of this interaction and its effects on the electrical transport properties of the SWNTs is therefore very important.

The lack of published data on an individual SWNT could be attributed to the technical difficulty in applying standard electron-beam lithography method for the fabrication of electrical terminals on an individual SWNT

* Correspondence: e_sadki@uaeu.ac.ae
[2]Physics Department, College of Science, United Arab Emirates University, Al Ain, Abu Dhabi, United Arab Emirates
Full list of author information is available at the end of the article on these substrates, as it is usually inseparable from the other SWNTs in the arrays.

In this letter, we present a method for the fabrication of electrical terminals on individual SWNTs aligned on an ST-quartz substrate and the measurement of their electrical transport properties from room temperature down to 2 K. The method consists of CVD synthesis of an individual SWNT from evaporated metal catalyst pad and shadow mask evaporation of metallic electrical contacts on the SWNT. The thickness and dimensions of the catalyst pad are optimized to yield on average one long and horizontally aligned single SWNT after CVD synthesis. In contrast to standard electron-beam lithography technique, this method has the advantage of not exposing the SWNTs to any electron beam irradiation or chemicals that are reported to damage or/and contaminate the SWNTs [16,17]. Furthermore, in order to minimize any damage or contamination of the SWNT before electrical properties measurements, scanning electron microscopy (SEM), Raman spectroscopy mapping, and atomic force microscopy (AFM) are performed only after all the electrical transport measurements are achieved. The electrical properties of individual SWNTs are measured using four-terminal method to minimize the effects of the contact resistance from the electrodes [18,19]. The results are compared with theory and discussed in connection with the strong interaction with the substrate.

## Methods

Figure 1 shows a schematic of the process of the synthesis of an individual SWNT and the fabrication of the electrical terminals on top of it. Titanium (Ti) film, with 2 μm thickness, is used as a shadow mask for the evaporation of cobalt catalyst pads. Catalyst pad patterns are milled in the titanium film using a focused ion beam (FIB) system (SMI9800SE, SII NanoTechnology Inc., Tokyo, Japan). The cobalt catalyst is evaporated through the titanium mask's patterns by electron beam (EB) evaporation, with a thickness of 2.0 nm, measured by a calibrated thickness monitor in the evaporator. After catalyst deposition, SWNTs are synthesized by thermal CVD method using a double zone furnace (ARF-30KC-W: Asahi Rika Corp., Chiba, Japan) equipped with a quartz tube of 27 mm in inner diameter. ST-cut quartz wafers (Hoffman Materials LLC., Carlisle, PA, USA), with crystallographic directions precisely defined within 0.08° by the manufacturer, are diced into rectangular substrates, with their longer side (length) exactly parallel to the $x$-direction of the crystal ([100] axis), which is the preferential growth direction of the SWNTs as reported by others [8,10,12]. The substrates are placed at the center of the downstream side of the furnace. A typical CVD process is as follows: Substrate is heated in a 200 sccm $O_2$ flow from room temperature to 900°C for 30 min to remove any carbon-based contaminations and allowed to settle for 5 min. Next, the $O_2$ flow is stopped and replaced by 300 sccm Ar flow for 10 min as a buffer gas after $O_2$ and before the introduction of hydrogen gas. Then, a 200 sccm $H_2$ gas is flown for 10 min to activate the cobalt catalyst film. Finally, the $H_2$ gas is co-flown with 300 sccm $CH_4$ gas for 15 min, which acts as the carbon source for SWNTs synthesis. Finally, the sample is left to cool down to room temperature in a continuous $H_2$ flow to prevent the oxidation of the SWNTs at high temperatures. The synthesized SWNTs were indeed confirmed to be parallel with the $x$-direction of the ST-cut quartz substrates as expected.

Electrodes on the SWNT are also fabricated using shadow mask evaporation technique. The metal masks are prepared by the same method as of that used for catalyst pattern. Palladium (Pd) is selected as the material of the electrodes because of its low contact resistance to SWNTs [20,21]. The Pd electrodes, with a thickness of 50 nm, are EB evaporated in a four-terminal configuration, with a typical distance of 4.0 μm between adjacent electrodes. The electrical properties of the SWNTs are measured from room temperature down to 2 K, using a physical properties measurement system (PPMS, Quantum Design Inc., San Diego, CA, USA) for the temperature control. Voltages of approximately ±1 V are applied by a voltage source (33220A, Agilent, Santa Clara, MA, USA) through a 10 MΩ resistance connected in series with the sample, and the voltage is measured across the inner electrodes on the sample by a voltmeter (Model 2000 Multimeter, Keithley, Cleveland, OH, USA).

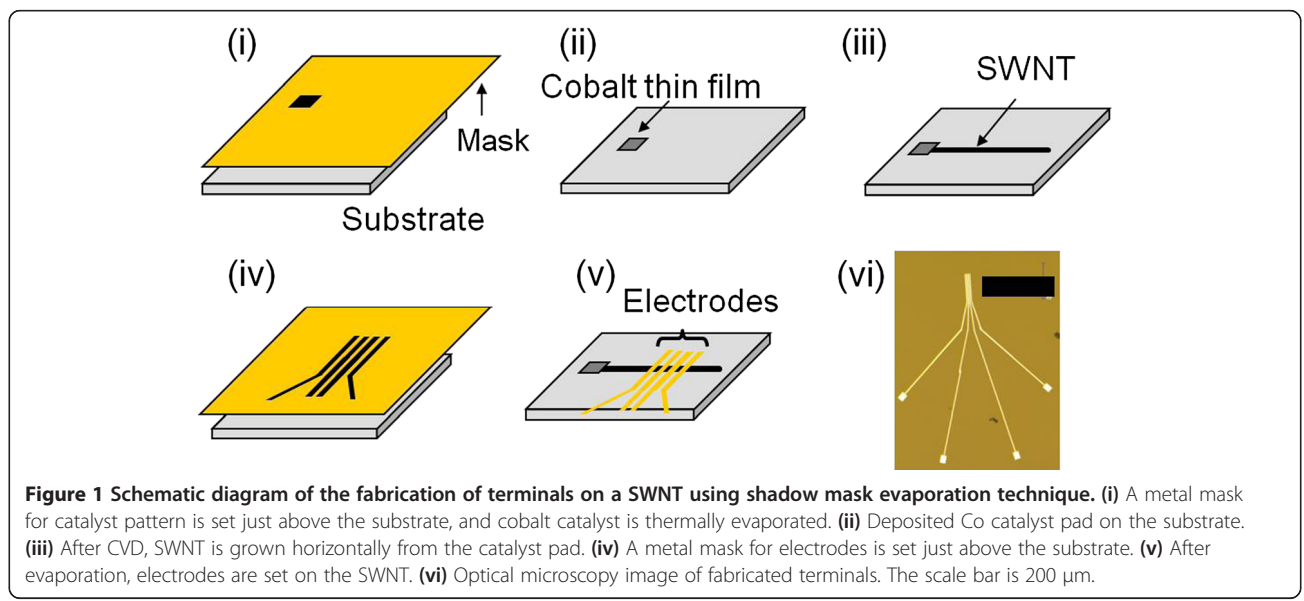

**Figure 1 Schematic diagram of the fabrication of terminals on a SWNT using shadow mask evaporation technique. (i)** A metal mask for catalyst pattern is set just above the substrate, and cobalt catalyst is thermally evaporated. **(ii)** Deposited Co catalyst pad on the substrate. **(iii)** After CVD, SWNT is grown horizontally from the catalyst pad. **(iv)** A metal mask for electrodes is set just above the substrate. **(v)** After evaporation, electrodes are set on the SWNT. **(vi)** Optical microscopy image of fabricated terminals. The scale bar is 200 μm.

For imaging and analytical characterization of SWNTs under the terminals, Raman spectral mapping (RAMAN-11, Nanophoton Corp., Osaka, Japan), AFM system (Nanocute, SII NanoTechnology Inc.), and SEM system (SMI9800SE, SII NanoTechnology Inc.) are used. Raman spectroscopy is performed with a laser of 532 nm in wavelength and spot size of 0.5 μm. AFM is conducted in cyclic contact AC mode.

## Results and discussion

In order to synthesize an individual and long SWNT for electrical characterization, the catalyst's pad dimensions are to be controlled accordingly. Figure 2a shows an SEM image of SWNTs synthesized from a catalyst pad of 100 × 10 μm in area. A lot of SWNTs are obtained in this case, with average lengths of more than 100 μm. On the other hand, as shown in Figure 2b, if a catalyst pad of 10 × 2 μm is used, only one or a few SWNTs are obtained, with typically the emergence of an individual SWNT of more than 100 μm in length. In order to precisely deposit the electrodes on a single SWNT, a specially designed substrate holder is used that keeps a fixed overlapping distance between the catalyst and electrode masks to within few microns resolution. Figure 2c shows deposited electrodes on a SWNT synthesized from the same pad's dimensions of 10 × 2 μm.

Figure 3a shows a typical AFM topography image of a SWNT between electrodes. It is noted that with the 2 nm thickness of the Co catalyst used, the obtained SWNTs have typical diameters of less than 1 nm. Figure 3b displays a Raman mapping image used to locate and confirm the presence of a single SWNT located between the electrodes. Figure 3c,d present the AFM thickness profiles of two nanotubes, denoted as SWNT1 and SWNT2, with estimated diameters of around 0.8 and 0.6 nm, respectively. It is noted that the measurement of SWNTs diameters by AFM is not accurate due to the roughness of the quartz substrate (typically 0.1 nm), as well as the interaction forces between the SWNTs and the substrate [11]. In order to precisely determine the diameter and chirality of our SWNTs, a study of the Raman spectrum of each SWNT is required [22]. Figure 3e shows the Raman spectra of the samples, where the G-band peaks are clearly observed for both SWNT1 and SWNT2. It is noted the absence of the D-band peaks from the spectra, which indicates that the synthesized SWNTs are nearly defect-free. However, the radial breathing mode (RBM) peaks were not observed in the spectra of both SWNTs. This indicates that the observed strong G-band signal from our individual SWNTs is from a resonance with the scattered photon, or $E_{laser} - E_{G\text{-band}} = E_{ii}$, where $E_{laser}$, $E_{G\text{-band}}$ (≈0.2 eV), and $E_{ii}$, are the laser's energy, the G-band phonons energy, and a SWNT's optical transition, respectively [22]. Applying the above condition on the Kataura plot (i.e., $E_{ii}$ vs diameter) [23], with $E_{laser}$ = 2.33 eV (532 nm wavelength) and a typical resonance window of 50 meV [22] points to two SWNTs satisfying the resonance condition with their $E_{22}$ optical transitions as shown in Figure 3f.

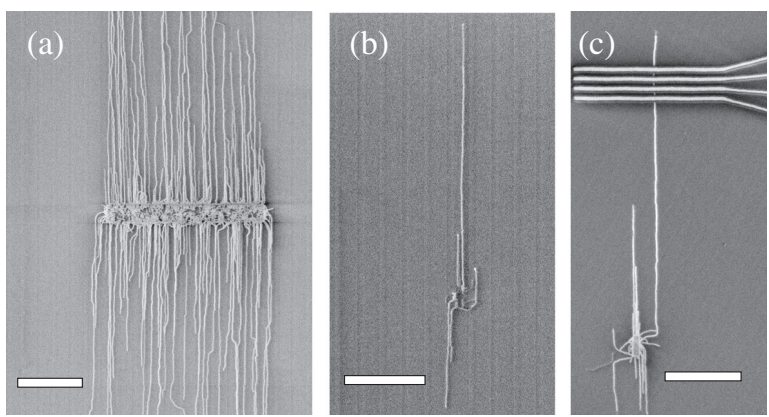

**Figure 2 SEM images of SWNTs synthesized from different catalyst pads.** Size of catalyst pad is 100 × 10 μm in (a), 10 × 2 μm in (b), and 10 × 2 μm in (c) with deposited electrodes. All scale bars are 40 μm.

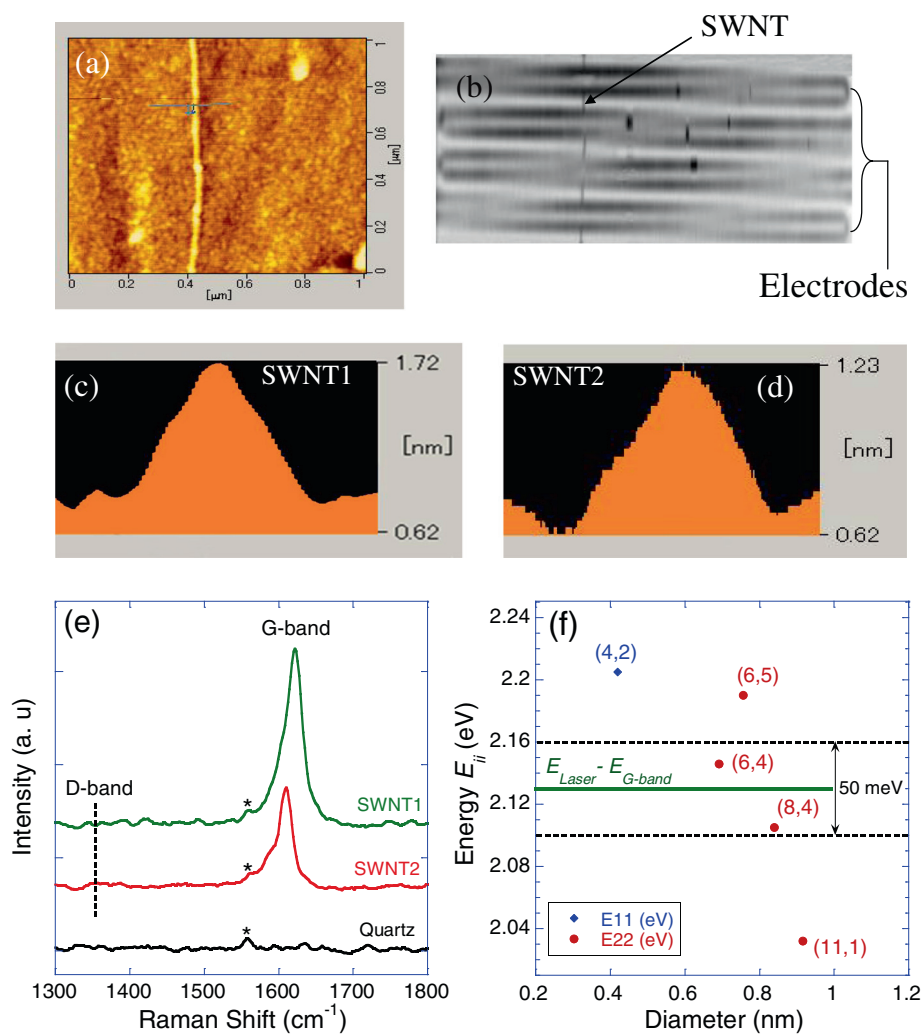

**Figure 3 AFM and Raman spectroscopy data analysis. (a)** AFM topography image of a SWNT. **(b)** Raman mapping image measured for a SWNT located between electrodes. **(c, d)** AFM topography profile for SWNT1 and SWNT2, respectively. **(e)** Raman spectra of the samples and the quartz substrate showing the G-band and the expected position of the D-band (dotted vertical line). The star marks show peaks attributed to the quartz substrate. **(f)** A Kataura plot of SWNTs optical energy transitions versus diameter showing the resonance region for the scattered photons (from the laser) with the G-band, with a resonance window of 50 meV. Two SWNTs fall within this region, namely (8,4) and (6,4), which correspond to SWNT1 and SWNT2, respectively.

Combining this result with the AFM data, it is clear that SWNT1 and SWNT2 correspond to the semiconducting nanotubes (8,4) and (6,4), respectively. This correspondence is achieved with a high degree of certitude as only two SWNTs felt within the Raman resonance condition of our experiment, and the theoretically calculated diameters of these SWNTs, namely 0.84 and 0.69 nm, for (8,4) and (6,4), respectively, are very close to the experimentally measured values by AFM.

From Figure 3e, it is observed that the G-band's peaks are located at frequencies 1621 and 1610 $cm^{-1}$, for SWNT1 and SWNT2, respectively. These values are significantly higher than the reported values of around 1590 $cm^{-1}$ for SWNTs on thermally grown silicon oxide substrates [24]. Similar up-shifts in the G-band have been observed for arrays of SWNTs aligned on ST-cut quartz and were attributed to the strong interaction between the SWNTs and the substrate [14,15]. However, our results provide a direct correlation between this up-shift in the G-band and the diameter and chirality of individual SWNTs. Since theoretical [22] and experimental results [25] show that the main peak of the G-band (i.e., the $G^+$ peak associated with longitudinal vibration of carbon atoms along the SWNT) is independent of the diameter and chirality for semiconducting SWNTs, it is concluded that the observed difference between SWNT1 and SWNT2 should be mainly due to the effect of the substrate. It is noted that the mechanism leading to the alignment of the SWNTs on ST-quartz substrates is attributed to a stronger and preferential interaction along the crystallographic direction [100] ($x$-axis) of the ST-quartz during CVD growth [26,27]. Based on a simple anisotropic Van der Waals interaction model between the SWNTs and the quartz substrate, Xiao et al. [26] predict an enhancement in this interaction with decreasing SWNT diameter. However, this is not in agreement with our results, where an increase in interaction (i.e., larger Raman up-shift) is observed with increasing diameter. On the other hand, assuming a shortened C-C bond (i.e., an increase in the force constant) along the SWNT's axis, experimental and theoretical works predict an up-shift in the G-band frequency [28,29], and that the effect is enhanced with increasing SWNT diameter and decreasing chiral angle [30,31]. This is indeed in agreement with our data if we assume that the interaction with the substrate causes a compression of the C-C bond along the SWNT's axis. It was stipulated that this interaction arises from a difference in the coefficient of thermal expansion between the SWNTs and quartz substrate when cooling down to room temperature after CVD growth [15]. However, the up-shift in the G-band was still observed even after transferring the SWNTs to another quartz substrate after CVD [14]. Therefore, the exact nature of the responsible mechanism for the G-band up-shift on these substrates is still unclear so far.

Figure 4 shows the results of the temperature dependence of the electrical resistance (normalized to its value at 300 K) of the two SWNTs measured with an electrical current of 10 nA. For SWNT1, the resistance decreases with decreasing temperature from room temperature down to about 120 K and then it increases by decreasing temperature down to 2 K. At the lowest temperature of 2 K, the resistance reaches about four times its room temperature value of 181 kΩ. On the other hand, the resistance of SWNT2 shows an increase with decreasing temperature from room temperature all the way down to 2 K. However, at 2 K, the normalized resistance reaches about 280 times its value at room temperature of 1.46 MΩ, which is more than 2 orders of magnitude higher than that in the case of SWNT1.

First, the values of the resistance at room temperature are considered. The intrinsic resistance of a SWNT in the diffusive regime (non-ballistic) can be estimated from the formula $R = R_c + R_Q(L/l + 1)$, where $R_c$, $R_Q = h/4e^2 \sim 6.45$ kΩ, $L$, and $l$ are the contact resistance between SWNT and the electrodes, the quantum resistance of a SWNT, the measured length of the SWNT, and the electron's mean free path, respectively [32]. By comparing the 2 and 4-terminal resistances of our samples, and using $L = 4$ μm (distance between the inner voltage terminals), $R_c$ and $l$ are estimated to be 8 and 19 kΩ, and 148 and 18 nm, for SWNT1 and SWNT2, respectively.

The deduced mean free paths for SWNT1 and SWNT2 at 300 K are within the range of reported values for SWNTs [18,33,34]. Nevertheless, it is very difficult to compare directly with our samples because most of the published electrical transport properties data either do not define the chirality of the measured SWNTs or it is about SWNTs with larger diameters than ours. In general, the SWNT's resistance at high temperatures is theoretically attributed to inelastic scattering between electrons and acoustic phonons within the SWNT [35]. However, the experimentally measured mean free paths of our SWNTs and others [18,33,34] are smaller by an order of magnitude than the theoretical calculations [35]. Recently, this discrepancy has been successfully addressed by introducing the effect of surface polar phonons (SPPs) from the substrate [36,37]. We speculate here that due to its narrower diameter, SWNT2 might be more susceptible to SPPs from the substrate, which enhance its room temperature resistance (i.e., shorter $l$) in comparison with SWNT1. It is noted from our results that the mechanisms defining the shift in the G-band and the electron's mean free path $l$ should be uncorrelated; otherwise, we would expect SWNT1 to have a shorter $l$. This is indeed in support of an extrinsic contribution of SPPs from the substrate than an intrinsic one from the SWNTs' own phonons. Further detailed studies on both contributions are therefore needed in the future.

Since SWNT1 is a semiconductor, the measured decrease of its resistance from room temperature down to about 120 K cannot be attributed to an intrinsic metallic property [38]. Based on the observed strong effect of the substrate on the G-band of SWNT1, we speculate that this metallic-like behavior could be originating from an interaction with the substrate that dominates at high temperature. Indeed, the expected semiconducting behavior of the resistance versus temperature is gradually recovered below around 120 K (Figure 4a). One possible indication for a semiconducting energy gap is a thermal activation dependence of the resistance versus temperature, i.e., in the form $R \sim \exp(U/k_B T)$, where $U$ and $k_B$ are an energy barrier and Boltzmann constant, respectively [39]. In order to explore this behavior, a plot of $Ln(R)$ versus $1/T$ is shown in Figure 4c, which could be very well fitted to the above activation formula from 60 K down to 5 K, with $U \sim 0.6$ meV. Assuming a standard semiconductor theory [39], this leads to a semiconducting energy gap of $E_g = 2U = 1.2$ meV. This value is about 2 orders of magnitude smaller than the expected and directly measured energy gap of 1.11 eV for SWNT1 [23]. This difference is not surprising as the simple activation formula above is used just as a qualitative guide, and the resistance versus temperature dependence of semiconducting SWNTs is very complex and there is no simple explicit formula in relation with $E_g$ [40]. A more accurate technique of extracting $E_g$ is from voltage-current measurements with a gating voltage [7]. However, this is not possible in our current experimental setup.

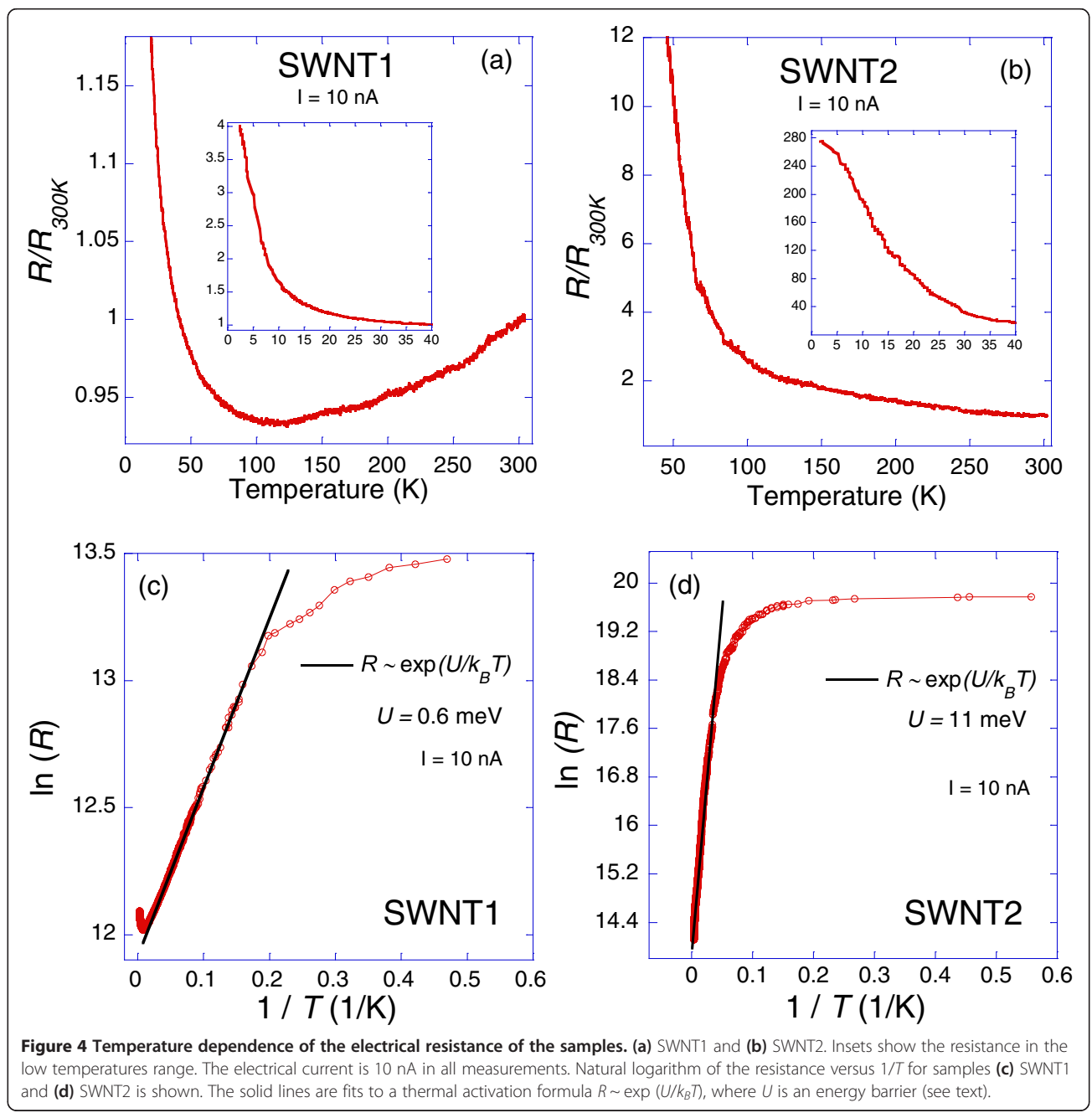

**Figure 4 Temperature dependence of the electrical resistance of the samples.** (a) SWNT1 and (b) SWNT2. Insets show the resistance in the low temperatures range. The electrical current is 10 nA in all measurements. Natural logarithm of the resistance versus $1/T$ for samples (c) SWNT1 and (d) SWNT2 is shown. The solid lines are fits to a thermal activation formula $R \sim \exp(U/k_B T)$, where $U$ is an energy barrier (see text).

The resistance of sample SWNT2 increases with decreasing temperature down to 2 K. In order to explore any thermal activation behavior, Figure 4d shows a plot of $Ln(R)$ versus $1/T$. The data from room temperature down to 20 K can be fitted very well with the activation formula, leading to an energy gap of $E_g = 2U = 22$ meV. This is in qualitative agreement with a semiconducting behavior in general but not quantitatively with $E_g = 1.42$ eV for SWNT2 [23], which is due to the same reasons explained before. It is noted that SWNT2 does not exhibit any decrease of $R$ with decreasing $T$ as observed for SWNT1. This could be due to a weaker effect from the substrate (less up-shift in G-band) than that of SWNT1 because of possibly the larger $E_g$ of SWNT2.

Since SWNTs are considered to be 1D systems, with strong electron–electron interaction, they are predicted to exhibit Tomonaga-Luttinger liquid (TLL) behavior at low temperatures [41-43]. Furthermore, SWNTs can act as a quantum dot between metal electrodes and hence show Coulomb blockade (CB) tunneling characteristics at sufficiently low temperatures [44-47]. Incidentally, both TLL and CB theories predict the same scaling laws: the resistance $R$ is proportional to $T^{-\alpha}$ when $eV << k_BT$ (low-bias regime) and to $V^{-\alpha}$ when $eV >> k_BT$ (high-bias regime), where $V$, $\alpha$, and $e$, are the voltage drop across the sample, a single scaling coefficient, and the charge of an electron, respectively [46]. In order to extract the values of $R$ in the two different regimes, current–voltage (IV) curves for both samples are measured at various temperatures as shown in Figure 5a,b. At high-bias voltages and low-bias voltage at high temperatures, the IV curves are basically linear with the current $I$ in both samples. However, at low bias and low temperatures, the IVs are not linear, especially in sample SWNT2. The origin of this curvature is discussed below.

First, for sample SWNT1, the low bias $R$ is extracted from the IV curves at $I = 1$ nA and plotted in a log-log graph versus temperature as shown in Figure 6a. The data fits well a power law above 30 K, with $\alpha \approx 0.1$. Note that $k_BT = 2.59$ meV $>> eV = 0.29$ meV at 30 K. This is in agreement with the regime of validity of the theory. Furthermore, the value of $\alpha \approx 0.1$ is in the same order as the reported values in the literature for SWNTs [41,46,47]. Next, $R$, in the high-bias regime, is extracted from the IV curves at $T = 2, 5,$ and 10 K and plotted in a log-log graph versus voltage as shown in Figure 6b. The low temperatures were chosen in order to be as close as possible to the condition $eV >> k_BT$ for this regime. For voltages $V$ higher than about 10 mV, the curve fits well a power law, with $\alpha \approx 0.1$. This is in very good agreement with the extracted value from $R$ versus $T$ in the other regime. Furthermore, knowing that $k_BT \approx 0.9$ meV at $T = 10$ K, the range of voltages where the power-law fit is found to hold (i.e., above 10 meV), indeed satisfies reasonably well the condition $eV >> k_BT$. The inset of Figure 6b shows that from 20 K and above, the resistance is essentially independent of the applied voltage, i.e., the IV curves are linear, which is exactly what was observed in Figure 5a. Hence, the behavior of SWNT1 is consistent with both LLD and CB theories with a scaling exponent $\alpha \approx 0.1$. First, it is noted that the extracted contact resistance, $R_c = 8$ kΩ, is higher than the quantum resistance $R_Q$, which satisfies a necessary condition for the occurrence of the CB [48]. Another theoretical condition for achieving CB is to have the charging energy $E_c$ of the SWNT higher than the thermal energy $k_BT$, with $E_c \approx 2.5$ meV/$L(\mu m)$ on $SiO_2$ [48]. This yields $E_c \approx 0.6$ meV for SWNT1, which requires a temperature $T < 7$ K for CB to occur. However, the scaling law is observed up to at least 10 K, which suggests that the observed scaling, at least above 7 K, could be indeed a TLL behavior. It is noted from Figure 6b that for bias voltages less than about 9 mV at 2 and 5 K, there is an increase in the resistance that could be attributed to enhanced CB effect with reducing bias voltages. This change in $R$ versus $V$ at low-bias voltages could be attributed to a crossover between the TLL and CB regimes [49]. Nevertheless, to experimentally confirm the CB effect, a gate voltage is required to modulate the SWNT's energy levels in order to possibly observe single electron tunneling as evidence for CB [37,40], which is beyond our current experimental setup.

The same TLL and CB scaling analysis is applied to sample SWNT2 as shown in Figure 6c,d. For $R$ vs $T$ plot, a fit to $T^{-\alpha}$ at high temperatures satisfying the low-bias condition $eV << k_BT$, yields an $\alpha \approx 0.5$. On the other hand, $R$ vs $V$ plot at the high-bias regime $eV >> k_BT$ leads to a power fit $V^{-\alpha}$, with $\alpha \approx 2$. Since the exponents from the two regimes are different, it is concluded that SWNT2 behavior is not consistent with TLL or CB. Figure 6d shows a dramatic increase in resistance at low

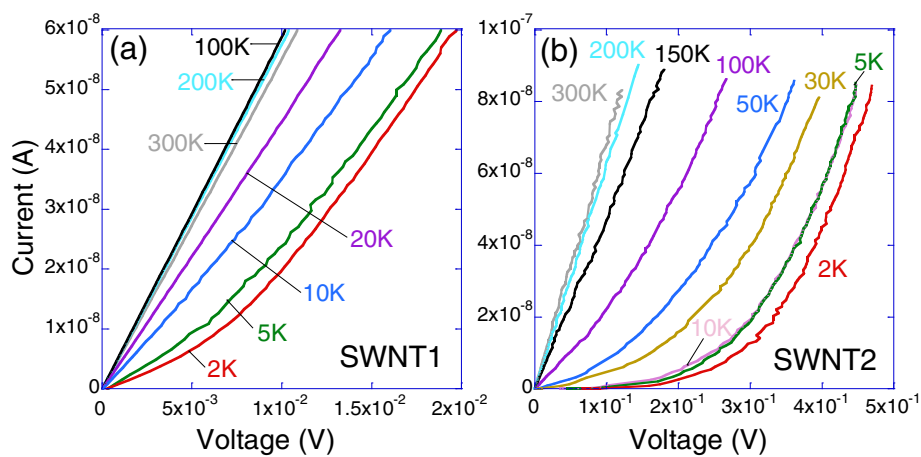

**Figure 5 Current–voltage (IV) curves.** For samples **(a)** SWNT1 and **(b)** SWNT2 measured at several temperatures from 300 to 2 K. Solid lines are guides to the eyes.

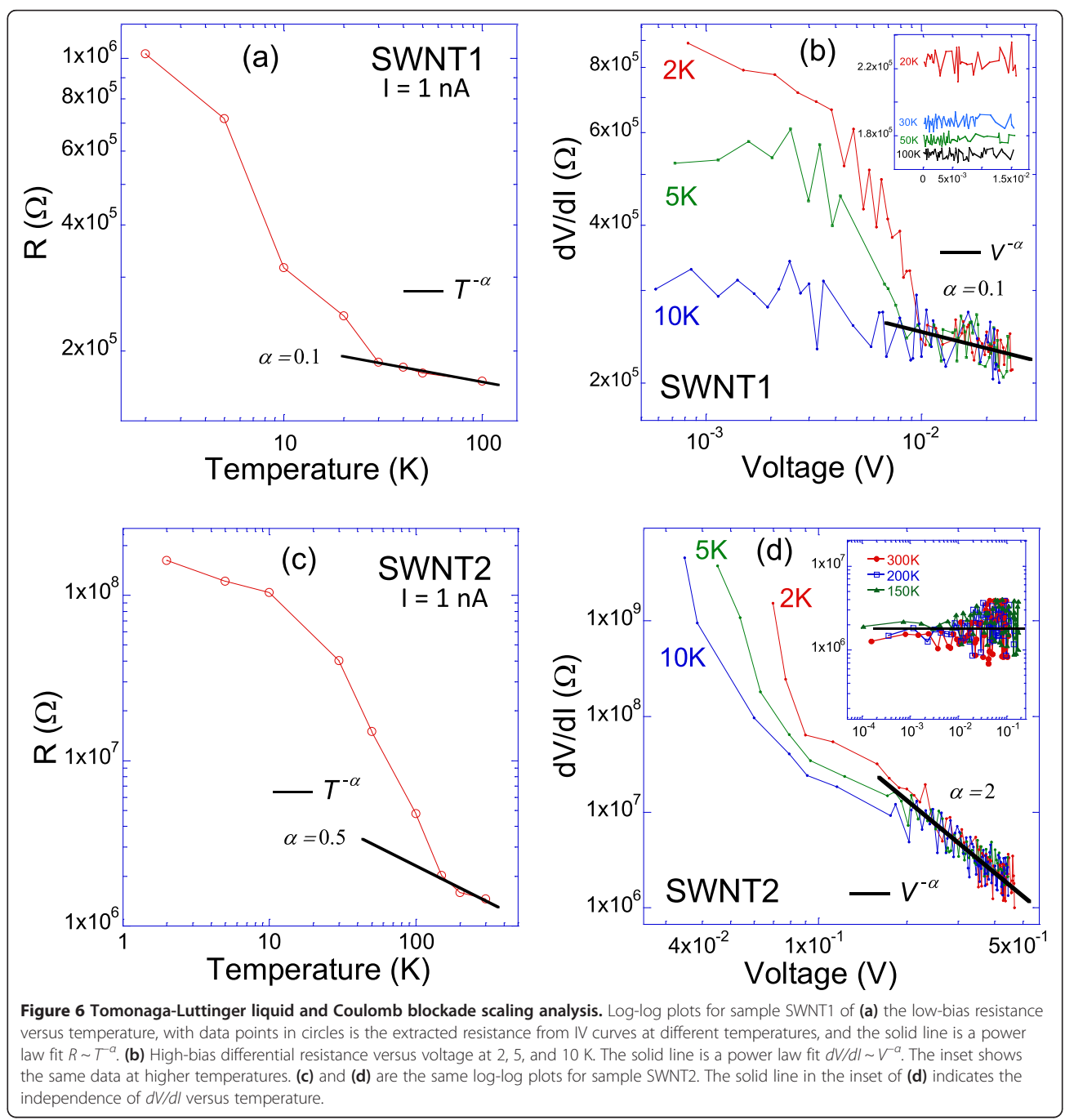

**Figure 6 Tomonaga-Luttinger liquid and Coulomb blockade scaling analysis.** Log-log plots for sample SWNT1 of **(a)** the low-bias resistance versus temperature, with data points in circles is the extracted resistance from IV curves at different temperatures, and the solid line is a power law fit $R \sim T^{-\alpha}$. **(b)** High-bias differential resistance versus voltage at 2, 5, and 10 K. The solid line is a power law fit $dV/dI \sim V^{-\alpha}$. The inset shows the same data at higher temperatures. **(c)** and **(d)** are the same log-log plots for sample SWNT2. The solid line in the inset of **(d)** indicates the independence of $dV/dI$ versus temperature.

bias for temperatures below or equal to 10 K. At higher temperatures, as shown in the inset of Figure 6d, the resistance is basically independent of the applied voltage, which is consistent with the linear IVs measured at higher temperature as shown in Figure 5b. The measured very high values of the resistance at low temperatures and low bias (in the order of GΩs) suggest the presence of an insulating state in this region. To explore this possibility, the current is plotted against voltage at the temperatures 2, 5, and 10 K, and low bias, as shown in Figure 7. Indeed, voltage thresholds separating a zero-resistance state (within the noise level of the measurements) and a conductive state at higher voltages are observed. The extracted values of these energy barriers are 82, 63, and 58 meV, for 2, 5, and 10 K, respectively, which are clearly much higher than the thermal energies $k_BT$ at these temperatures. Such insulating state in individual SWNTs have been observed by some other groups [50,51]. An energy barrier of 600 meV has been observed from an IV curve at 4 K for a semiconducting

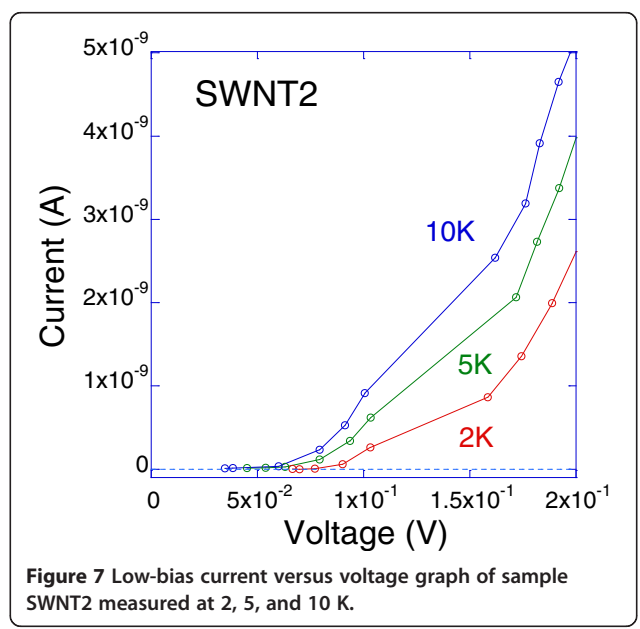

Figure 7 Low-bias current versus voltage graph of sample SWNT2 measured at 2, 5, and 10 K.

SWNT and was attributed to the intrinsic gap of the SWNT [50]. This value is less than that of SWNT2 (1.42 eV) but still in the same order of magnitude for a qualitative comparison. In the 2-point measurements of Zhou et al. [50], the contact resistance is included in their IVs, which might induce a barrier due to metal–semiconductor-metal junction effects. This is excluded or at least minimized in our 4-point measurements, as the contact resistance is subtracted in our configuration. Furthermore, the estimated contact resistance $R_c$ of SWNT2 is less than $3R_Q$, which is reasonably too small to be considered as invasive or to induce a significant contact barrier [40]. Interestingly, from measurements on suspended (no substrate effects) and 'ultraclean' metallic SWNTs, a Mott insulating state was reported, with energy gaps between 10 and 100 meV [51]. Specifically, for SWNTs with diameters similar to SWNT2, the energy barrier was between 70 and 80 meV, which is in good agreement with the measured barriers for SWNT2. However, to explore the nature of the insulating state in SWNT2, gating experiments are needed, which is again beyond the scope of this letter.

Finally, the appearance of completely different properties for SWNT1 (TLL/CB) and SWNT2 (transition to an insulating state) at low temperatures and their relation with the observed strong interaction with the quartz substrate is currently not understood. Further theoretical and experimental efforts are underway to elucidate these effects.

## Conclusions

In conclusion, a method is introduced to isolate and measure the electrical properties of individual SWNTs aligned on an ST-cut quartz substrate, from room temperature down to 2 K. The diameter and chirality of the measured SWNTs are accurately defined from resonant Raman spectroscopy and AFM. A significant up-shift in the G-band of the Raman spectra of the SWNTs is observed, which increases with increasing SWNTs diameter and indicates a strong interaction with the quartz substrate. A semiconducting SWNT (diameter 0.84 nm) shows Tomonaga-Luttinger liquid and Coulomb blockade behaviors at low temperatures. Another semiconducting SWNT (diameter of 0.68 nm) exhibits a transition from the semiconducting state to an insulating state at low temperatures. These results elucidate some of the electrical transport properties of SWNTs on ST-cut quartz substrates, which can be useful for prospective device applications.

**Competing interests**
The authors declare that they have no competing interests.

**Authors' contributions**
TW synthesized, characterized, and interpreted the data of the SWNTs, as well as drafted the initial version of the manuscript. ESS had the original idea of the project, contributed to the experimental setup, interpreted the data, and drafted the final manuscript with TW. TY contributed with the experimental setup and transport measurements of the SWNTs. YT coordinated the project and supervised TW. All authors read and approved the final manuscript.

**Acknowledgements**
This study was supported by Nano-Integration Foundry (NIMS) in 'Nanotechnology Platform Project' operated by the Ministry of Education, Culture, Sports, Science and Technology (MEXT), Japan. ESS would like to acknowledge the support and hospitality of NIMS during his visit as a Guest Researcher.

**Author details**
[1]National Institute for Materials Science, 1-2-1, Sengen, Tsukuba, Ibaraki 305-0047, Japan. [2]Physics Department, College of Science, United Arab Emirates University, Al Ain, Abu Dhabi, United Arab Emirates.